\documentclass{article}
\usepackage{epic}
\usepackage{eepic}

\long\def\nop#1{}

\def\comment{\edef\cps{\the\parskip} \parskip=0.5cm \begingroup \tt}
\def\endcomment{\endgroup \vskip 0.5cm \parskip=\cps}

\expandafter\ifx\csname shortcite\endcsname\relax

\fi

\newbox\current

\long\def\plframebox#1{
\setbox\current\vbox{#1}		

\vbox to \ht\current {\hrule\vss
\hbox to \wd\current {%
\vrule \hss\box\current\hss \vrule}
\vss\hrule }
}

\long\def\eatpar#1{%
\ifx#1\par                      
\let\nextmove=\eatpar           
\else
\let\nextmove=#1
\fi
\noexpand\nextmove
}

\def\modifymargins#1#2{
\newdimen\addtoh
\newdimen\addtow
\addtoh=#1
\addtow=#2

\advance\topmargin by -\addtoh
\multiply\addtoh by 2
\advance\textheight by \addtoh

\advance\oddsidemargin by -\addtow
\advance\evensidemargin by -\addtow
\multiply\addtow by 2
\advance\textwidth by \addtow
}

\begingroup
\catcode`\~=11
\gdef\centertilde#1{\lower #1pt\hbox{~}}
\endgroup

\newcount\currenttime
\newcount\hour
\newcount\minute

\def\printtime{%
\currenttime=\time
\hour=\currenttime
\divide\hour by 60
\minute=-\hour
\multiply\minute by 60
\advance\minute by \currenttime
\the\hour:\ifnum\minute<10 0\fi\the\minute
}

\begingroup
\makeatletter
\global\let\@@date=\@date
\gdef\@date{\@@date\ --- \printtime}
\endgroup

\def\oggi{\number\day\space 
\ifcase\month\or
Gennaio\or Febbraio\or Marzo\or Aprile\or Maggio\or Giugno\or
Luglio\or Agosto\or Settembre\or Ottobre\or Novembre\or Dicembre\fi
\space \number\year}

\newcounter{rmexample}

\def\proof{\noindent {\sl Proof.\ \ }}

\def\qed{\hfill{\boxit{}}
  \ifdim\lastskip<\medskipamount \removelastskip\penalty55\medskip\fi}
\def\qedn#1{\hfill{\boxit{}$_#1$}
  \ifdim\lastskip<\medskipamount \removelastskip\penalty55\medskip\fi}
\long\def\boxit#1{\vbox{\hrule\hbox{\vrule\kern3pt
                  \vbox{\kern3pt#1\kern3pt}\kern3pt\vrule}\hrule}}

  \def\D{{\cal D}}

\def\ie{i.e.}
\def\eg{e.g.}
\def\wrt{w.r.t.}

\def\mod{M\!od}

\def\var{V\!ar}

\def\true{{\sf true}}
\def\false{{\sf false}}

\def\C{{\rm C}}

\def\p{{\rm P}}
\def\np{{\rm NP}}
\def\conp{{\rm coNP}}
\def\Dp{${\rm D}^p$}

\def\D#1{\mbox{$\Delta^p_{#1}$}}
\def\Dlog#1{\mbox{$\Delta^p_{#1}[\log n]$}}

\def\pp{{\rm PP}}

\def\nppp{$\np^\pp$}

\def\pspace{{\rm PSPACE}}

\def\profont{\sf}

\def\x3c{{\profont x3c}}

\def\possnewtheorem#1#2{
\expandafter\ifx\csname #1\endcsname\relax
\newtheorem{#1}{#2}
\fi
}

\def\possnewtheoremthree#1[#2]#3{
\expandafter\ifx\csname #1\endcsname\relax
\newtheorem{#1}[#2]{#3}
\fi
}

\possnewtheorem{theorem}{Theorem}
\possnewtheorem{corollary}{Corollary}
\possnewtheorem{lemma}{Lemma}
\possnewtheoremthree{proposition}[theorem]{Proposition}
\possnewtheorem{definition}{Definition}
\possnewtheorem{question}{Question}
\possnewtheorem{example}{Example}
\possnewtheorem{property}{Property}
\possnewtheorem{assumption}{Assumption}
\possnewtheorem{conjecture}{Conjecture}
\newenvironment{theorem*}[1]{{\noindent \bf Theorem~#1}\begin{it}}{\end{it}\

}

\long\def\comment#1\endcomment{}

\def\paritysat{{\sc parity(sat)}}

\title{Complexity Results on DPLL and Resolution}
\author{Paolo Liberatore\\
\normalsize Universit\`a di Roma ``La Sapienza''\\
\normalsize Dipartimento di Informatica e Sistemistica,\\
\normalsize Universit\`a di Roma ``La Sapienza'',\\
\normalsize Via Salaria 113, 00198, Rome, Italy.\\
\normalsize Email: \tt paolo@liberatore.org
}

\begin{document}

\maketitle

\begin{abstract}

DPLL and resolution are two popular methods for solving the
problem of propositional satisfiability. Rather than
algorithms, they are families of algorithms, as their
behavior depend on some choices they face during execution:
DPLL depends on the choice of the literal to branch on;
resolution depends on the choice of the pair of clauses to
resolve at each step. The complexity of making the optimal
choice is analyzed in this paper. Extending previous
results, we prove that choosing the optimal literal to
branch on in DPLL is \Dlog{2}-hard, and becomes
$\np^\pp$-hard if branching is only allowed on a subset of
variables. Optimal choice in regular resolution is both
\np-hard and \conp-hard. The problem of determining the size
of the optimal proofs is also analyzed: it is \conp-hard for
DPLL, and \Dlog{2}-hard if a conjecture we make is true.
This problem is \conp-hard for regular resolution.

\end{abstract}

 %

\section{Introduction}

\begin{sloppy}

Several algorithms for solving the problem of propositional
satisfiability exist. Among the fastest complete ones are
DPLL \cite{davi-loge-love-62,davi-putn-60} and resolution
\cite{robi-65}. Both of them depend on a specific choice to
make during execution. DPLL is a form of backtracking, and
therefore depends on how the branching variable is chosen.
Resolution runs by iteratively combining (resolving) two
clauses to obtain a consequence of them, until contradiction
is reached or any other clause that can be generated is
subsumed by one already generated. The choice of the
variable to branch on and the choice of the clauses to
combine (resolve) are crucial to efficiency. Formally, both
DPLL and resolution are {\em families} of algorithms: each
algorithm corresponds to a specific way for making the
choices, and can be very different to the other ones of the
same family as for its efficiency. Making the right choice
is therefore very important for ensuring efficiency. In this
paper, we show that this problem is \Dlog{2}-hard for DPLL
and backtracking, it becomes $\np^\pp$-hard for
restricted-branching DPLL (the variant of DPLL in which
branching is only allowed on a subset of the variables), and
is \conp-hard for regular resolution.

\end{sloppy}

A related problem is that of checking the size of the
optimal proofs. Indeed, while satisfiability of
propositional formulae can be proved with a very short
``certificate'' (the satisfying assignment),
unsatisfiability (probably) requires exponential proofs, in
general. Checking the size of the optimal proofs of an
unsatisfiable formula is important for at least two reasons:
the unsatisfiability proof may be necessary (for example, it
is the input of another program), and its size being too
large makes it practically useless; moreover, if the size of
optimal proofs can be checked efficiently, we may decide to
use incomplete methods to solve the satisfiability problem
if the size of the optimal proofs is too large (to be more
precise, if we can decide whether the formula is either
satisfiable or has a short proof efficiently, then we can
check the size of the proof to choose the algorithm.) We
prove that the problem of proof size is \conp-hard for DPLL
and backtracking (\Dlog{2}-hard if a conjecture we make is
true), $\np^\pp$-hard for restricted-branching DPLL, and
\conp-hard for regular resolution.

While the problem of making the right choice is the one that
has been more studied in practice \cite{li-anbu-97,li-2000},
it is the proofs size one that has been more investigated
from the point of view of computational complexity, perhaps
because it is also related to the question of relative
efficiency of proof methods. The proof size problem is known
to be \np-complete \cite{iwam-miya-95,iwam-97,alek-etal-01}.
Membership to \np\  only holds if the maximal size of proofs
is represented in unary notation, while the results
presented in this paper hold for the binary notation. The
meaning of the difference between the binary or unary
representation is discussed in the Conclusions.

A problem that is related to that of the optimal choice is
that of automatizability, which has recently received
attention \cite{beam-etal-02,bone-etal-99,bone-etal-00-c}.
Roughly speaking, a complete satisfiability algorithm is
automatizable if its running time is polynomial in the size
of the optimal proofs (and, therefore, it generates an
almost-optimal proof.) Some recent results have shown that,
in spite of some partially positive and unconditioned
results \cite{beam-etal-02,alek-razb-02}, resolution is not
automatizable in general \cite{alek-razb-01}.

The paper is organized as follows: in
Section~\ref{preliminaries}, we give the needed definitions
and some preliminary results; in Section~\ref{sec-dpll-back}
we show the complexity of making the optimal choice in DPLL
and backtracking; in Section~\ref{restricted} we analyze the
restricted-branching version of DPLL; in
Section~\ref{sec-resolution} we consider the complexity of
making the optimal choice in resolution. Discussion of the
results and comparison with related work is given in
Section~\ref{sec-conclusions}.

 %

\section{Preliminaries}
\label{preliminaries}

In this paper we analyze solvers for propositional
satisfiability. We assume that formulae are in CNF, \ie,
they are sets of clauses, each clause being a disjunction of
literals. For example, $\{ x_1 \vee x_2, \neg x_3 \}$ is the
set composed by the two clauses $ x_1 \vee x_2$ and $\neg
x_3$. We use the following notation:
$
l \vee F = \{ l \vee \gamma ~|~ \gamma \in F \}
$, where $F$ is a formula and $l$ is a literal.

A propositional interpretation is a mapping from the set of
variables to the set $\{\true, \false\}$. We denote an
interpretation by the set of literals containing $x$ or
$\neg x$ depending on whether $x$ is assigned to $\true$ or
$\false$. This notation is also used to represent partial
interpretations: if the set neither contains $x$ nor $\neg
x$, then the variable $x$ is unassigned. We denote the set
of models (satisfying assignments) of a formula $F$ by
$\mod(F)$. We denote the cardinality of a set $S$ by $|S|$;
therefore, $|\mod(F)|$ is the number of models of $F$.

If $F$ is a formula and $I$ is a partial interpretation,
$F|I$ denotes the formula obtained by replacing each
variable that is evaluated by $I$ with its value in $F$ and
then simplifying the formula. The resulting formula only
contains variables that $I$ leaves unassigned.

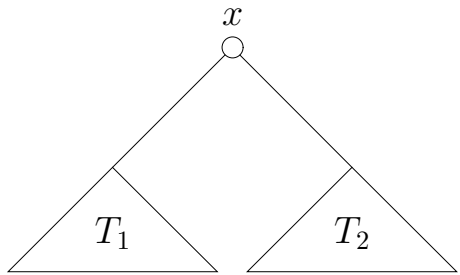
\begin{figure}[ht]
\centering
\setlength{\unitlength}{0.00087489in}
\begingroup\makeatletter\ifx\SetFigFont\undefined%
\gdef\SetFigFont#1#2#3#4#5{%
  \reset@font\fontsize{#1}{#2pt}%
  \fontfamily{#3}\fontseries{#4}\fontshape{#5}%
  \selectfont}%
\fi\endgroup%
{\renewcommand{\dashlinestretch}{30}
\begin{picture}(2724,1656)(0,-10)
\put(1362,1362){\ellipse{128}{128}}
\path(1317,1317)(642,642)
\path(1407,1317)(2082,642)
\path(2082,642)(1452,12)(2712,12)(2082,642)
\path(642,642)(12,12)(1272,12)(642,642)
\put(642,147){\makebox(0,0)[b]{{\SetFigFont{14}{16.8}{\familydefault}{\mddefault}{\updefault}$T_1$}}}
\put(2082,147){\makebox(0,0)[b]{{\SetFigFont{14}{16.8}{\familydefault}{\mddefault}{\updefault}$T_2$}}}
\put(1362,1497){\makebox(0,0)[b]{{\SetFigFont{14}{16.8}{\familydefault}{\mddefault}{\updefault}$x$}}}
\end{picture}
}
 %
%
\caption{Graphical representation of the tree $(x~T_1~T_2)$.}
\label{fig-a-tree}
\end{figure}

Proofs of unsatisfiability as built by the DPLL and
backtracking algorithms are binary trees whose nodes are
variables. We use the recursive definition of binary trees:
a tree is either empty, or is a triple composed of a node
and two trees. Trees will be represented either graphically
or in parenthetic notation. In the parenthetic notation,
$()$ denotes the empty tree, and $(x ~ T_1 ~ T_2)$ denotes
the nonempty tree whose label of the root is $x$ and whose
left and right subtrees are $T_1$ and $T_2$, respectively. A
{\em leaf} is a tree composed of a node and two empty
subtrees, \eg, $(x~()~())$. The {\em size} of a tree is the
number of nodes it contains. In some points, we write ``the
empty subtrees of $T$'' to indicate any empty tree that is
contained in $T$ or in any of its subtrees. By an inductive
argument, the number of empty subtrees of a tree is equal to
the number of its nodes plus one. The sentence ``replace
every empty subtree of $T_1$ with $T_2$'' has the obvious
meaning. The tree that is denoted by $(x~T_1~T_2)$ in
parenthetic notation is graphically represented as in
Figure~\ref{fig-a-tree}.

This graphical representation justifies the use of terms
such as ``above'', ``below'', etc., to refer to the relative
position of the nodes in the tree. If $A$ is a tree or a
formula, we denote by $\var(A)$ the set of variables it
contains.

\subsection{Backtracking and the DPLL Algorithm}

The DPLL algorithm is a backtracking algorithm working in
the search space of the partial models, enhanced by three
rules. The backtracking algorithm can be described as
follows: choose a variable among the unassigned ones, and
recursively execute the algorithm on the two subproblems
that result from setting the value of the variable to
$\false$ and $\true$. The base case of recursion is when
either all literals of a clause are falsified (the formula
is unsatisfiable), or when every clause contains at least
one true literal (the formula is satisfiable.) The whole
formula is satisfiable if and only if either recursive call
returns that the sub-formula is satisfiable.
Unsatisfiability therefore leads to a tree of recursive
calls, in which unsatisfiability is proved in each leaf.
This tree is called the search tree of the formula. A
formula can have several search trees, each one
corresponding to a different way of choosing the variables
to branch on.

\begin{definition}

A backtracking search tree (BST) of a formula $F$ is:

\begin{enumerate}

\item the empty tree $()$ if $F$ contains an empty clause (a
contradiction);

\item a non-empty tree $(x~T_1~T_2)$ otherwise, where $x \in
\var(F)$, and $T_1$ and $T_2$ are BSTs of $F|\{\neg x\}$ and
$F|\{x\}$, respectively.

\end{enumerate}

\end{definition}

DPLL~\cite{davi-loge-love-62,davi-putn-60} enhances the
backtracking procedure with three rules: {\em unit
propagation} consists in setting the value of a variable
whenever all other variables of a clause are false; the {\em
monotone literal rule} sets the value of variables that
appear with the same sign in the whole formula; {\em clause
subsumption} consists in removing clauses that are subsumed
by other ones. Clause subsumption is not used in modern
implementations of DPLL, and we therefore disregard it. The
search trees of DPLL are similar to those of backtracking.

\begin{definition}

Let $D(F)$ denotes the formula obtained from $F$ by applying
the unit propagation and monotone literal rules. A DPLL
search tree (DST) of a formula $F$ is:

\begin{enumerate}

\item the empty tree $()$ if $D(F)$ contains an empty
clause (a contradiction);

\item a non-empty tree $(x~T_1~T_2)$ otherwise, where $x \in
\var(D(F))$, and $T_1$ and $T_2$ are DSTs of $D(F|\{\neg
x\})$ and $D(F|\{x\})$, respectively.

\end{enumerate}

\end{definition}

An optimal (backtracking or DPLL) search tree for a formula
$F$ is a minimal-size search tree of $F$ (the size of a tree
is the number of nodes it contains.) A variable is an
optimal branching variable for $F$ if it is the root of an
optimal search tree of~$F$.

In general, the BSTs and the DSTs of a formula are not the
same. For example, $(x_1 (x_2()()) (x_2()()))$ and
$(x_1()())$ are a BST and a DST of $\{ \neg x_1 \vee x_2,
x_1 \vee \neg x_2 \}$, respectively, but not vice versa.
Nevertheless, a correspondence between BSTs and DSTs can be
established: for each formula $F$, we can build a new one
$G$ in such a way the DSTs of $G$ can be converted into BSTs
of $F$.

\begin{lemma}[{\cite[Lemma~1]{libe-00}}]
\label{dpll-back}

Let $F$ be a formula over variables $\{x_1,\ldots,x_n\}$,
$F'$ be obtained from $F$ by replacing every positive
literal $x_i$ with $x_i \vee y_i$ and every negative literal
$\neg x_i$ with $\neg x_i \vee \neg y_i$, and $G$ be defined
as follows:

\[
G =
\{ x_i \vee \neg y_i, \neg x_i \vee y_i ~|~ 1 \leq i \leq n \}
\cup
F'
\]

\noindent Every BST of $F$ is a DST of $G$, and every DST of
$G$ can be transformed into a BST of $F$ by replacing each
node labeled with $y_i$ with a node labeled with $x_i$ and
swapping its two subtrees.

\end{lemma}

This theorem shows that backtracking can be ``simulated'' by
DPLL: we can reproduce the backtracking behavior (and,
therefore, its search trees) using DPLL. This result is used
to prove the hardness of some problems about DPLL from the
corresponding ones about backtracking. For example, the
optimal BST size of a formula $F$ is equal to the size of
the optimal DST of the corresponding formula $G$; since the
translation from $F$ to $G$ can be done in polynomial time,
the problem of finding the optimal DST size for DPLL is at
least as hard as the corresponding problem for backtracking.

We denote by $s(F)$ the size of the optimal search tree of
the set of clauses $F$ if it is unsatisfiable, and $\infty$
otherwise. Whether we consider the backtracking or the DPLL
search trees can be inferred from the context.

\subsection{Resolution}

Resolution is a proof method based on the following rule:
if $\gamma \vee x$ and $\delta \vee \neg x$ are two clauses,
then $\gamma \vee \delta$ is a consequence of them. This
step of generating a new clause from two ones is called the
resolution of the two clauses; the generated clause is
called the resolvent. Satisfiability can be established
using the fact that this rule is complete \cite{robi-65}:
if a set of clauses is unsatisfiable, then the empty
clause (the clause with no literals) can be generated by
repeating the application of the resolution rule. Efficiency
clearly depends on how we choose, at each step, the pair of
clauses to resolve. 

The clauses generated to prove unsatisfiability can be
arranged into a DAG, in which the parent of two clauses is
their resolvent. The root of this DAG is the empty clause,
and the leaves are the clauses of the original set. If no
path from the root to a leaf contains two times the
resolution of the same variable, the resolution is called
regular. Regular resolution is the process of checking
unsatisfiability using a regular resolution proof.

 %

\subsection{Complexity of What?}

The results in this paper are about the complexity of making
choices in the DPLL and resolution procedures. Namely, we
consider the problem of making the first choice optimally.
For DPLL and backtracking, the problem is defined as
follows.

\begin{description}

\item[{\sc Name:}] Optimal branching variable (OBV)

\item[{\sc Instance:}] A formula $F$ and a variable $x$;

\item[{\sc Question:}] Is $x$ the root of an optimal
search tree of $F$?

\end{description}

This is a decision problem, that is, its solution is either
``yes'' or ``no''. The related problem of finding an optimal
branching variable can be solved by checking optimality for
all variables of the formula.

A related problem is that of finding the size of the optimal
search trees of a given formula. The formal definition is as
follows.

\begin{description}

\item[{\sc Name:}] Optimal tree size (OTS)

\item[{\sc Instance:}] A formula $F$ and an integer $k$ in
binary notation;

\item[{\sc Question:}] Does $F$ have a search tree of size
bounded by $k$?

\end{description}

The variant of OTS where $k$ is in unary notation has
already been investigated and proved \np\ complete
\cite{iwam-miya-95,buss-95,iwam-97} (the assumption that $k$
is in unary is necessary to prove the membership to \np.)
Assuming that $k$ is in unary notation means that the
complexity of the problem is not measured \wrt\  the size of
the formula $F$, but rather w.r.t.\  the size of the proof
we are looking for, which can be exponentially larger. We
assume that $k$ is in binary notation instead. The
difference between the binary and unary notation is
discussed in the Conclusions.

About resolution, the problem we consider is whether the
resolution of a pair of clauses is at the leaf level of an
optimal regular resolution proof. This is again the problem
of making the first choice optimally when using regular
resolution. Formally, the decision problem we analyze is the
following one.

\begin{description}

\item[{\sc Name:}] Optimal resolution pair (ORP)

\item[{\sc Instance:}] A formula $F$ and two of its clauses
$\gamma$ and $\delta$;

\item[{\sc Question:}] Is there an optimal regular
resolution proof of $F$ that contains the resolution of the
leaves $\gamma$ and $\delta$?

\end{description}

We ask whether two clauses are brother leaves of a regular
resolution proof (a DAG), while for DPLL the question is
about the root of a tree. This difference is due to the way
these procedures build their proofs: DPLL starts from the
root, resolution starts from the leaves. In both cases, the
problem we consider is that of making the first choice
optimally.

The problems we have presented in this section will be
characterized in terms of complexity classes. Some of the
classes we use are not well known as \np\  and \conp, so we
briefly recall their definition. A machine that works with
an oracle for the class \C\  is a model of computation that
can solve a problem in \C\  in a unit of time. The class
\D{2} is the class of problems that can be solved by a
machine that works in polynomial time with an oracle in \np.
The class \Dlog{2} is similar, but the oracle can only be
queried at most a logarithmic number of times. The class
\pp\ contains all problems that can be reduced to that of
deciding whether a propositional formula is satisfied by at
least half of the possible truth assignment of its
variables. The class \Dp\ contains all problems that can be
expressed as $L_1 \cap L_2$, where $L_1$ and $L_2$ are in
\np\  and \conp, respectively.

 %

\subsection{Combining Sets of Clauses}

In this section, we prove some general results about BSTs:
first, we show a formula whose optimal BSTs have size within
a given range; second, we show how to combine two sets of
clauses having some control on the size of the optimal BSTs
of the result. The first result is simply an adaptation of a
result by Urquhart~\cite{urqu-87} to backtracking.

\begin{lemma}
\label{exp}

For any given square number $m$, one can find, in time
polynomial in the value of $m$, a set of clauses $H_m$ over
$m$ variables whose optimal BSTs have size between $2^{cm}$
and $2^{m}$, where $c$ is a constant ($0 < c < 1$).

\end{lemma}

Since the algorithm that finds $H_m$ from $m$ runs in time
polynomial in $m$, the produced output $H_m$ is necessarily
of size polynomial in the value of $m$.

The first method we use for combining two sets of clauses is
the union. If two sets do not share variables, the optimal
BSTs of their union are simple to determine.

\begin{lemma}[\cite{libe-00}, Lemma~3]
\label{min}

If $F$ and $H$ are two sets of clauses not sharing any
variables, and $F \cup H$ is unsatisfiable, the optimal BSTs
of $F \cup H$ are optimal BSTs of one of the unsatisfiable
sets between $F$ and $H$.

\end{lemma}

In other words, if either $F$ or $H$ is satisfiable, the
optimal BSTs of $F \cup H$ are the optimal BSTs of the other
formula. If both $F$ and $H$ are unsatisfiable, the optimal
BSTs of $F \cup H$ are the smallest among the BSTs of $F$
and $H$.

The second way for combining two sets of clauses is what we
call ``addition''. This name has been chosen because the
size of the optimal BSTs of the combination is the sum of
the size of the optimal BSTs of the components.

\begin{definition}

The {\em sum} of two sets of clauses $F$ and $H$ is:

\[
F +_{x} H = (F \vee x) \cup (H \vee \neg x)
\]

\noindent where $x$ is a new variable not contained
in any of the two sets. When we do not care about the
name of the new variable, we omit it and write $F+H$.

\end{definition}

We remark that, if either $F$ or $H$ is satisfiable, their
addition is satisfiable.

\begin{lemma}
\label{add-variable}

Let $F$ and $H$ be two sets of clauses built over two
disjoint sets of variables, and let $x$ be a variable not
contained in them. If both $F$ and $H$ are unsatisfiable,
$x$ is an optimal backtracking branching literal for $F
+_{x} H$.

\end{lemma}

\proof We prove this lemma by induction over the
total number of variables of $F$ and $H$. The base case is
true: if neither $F$ nor $H$ contain any variable, then they
can be unsatisfiable only if they both contain the empty
clause (the contradiction). Their sum is therefore $\{ x,
\neg x\}$. For the induction case, if the statement of the
theorem holds for $n \geq 0$, and $T$ is an optimal BST of
$F +_x H$, then either $x$ is its root, or it is the root of
both its subtrees (because of the induction hypothesis). In
the second case, the tree can be reshaped to have $x$ in the
root.~\qed

The optimal BSTs of $F +_{x} H$ having $x$ in the root have,
as subtrees, optimal BSTs of $F$ and $H$. As a result, if
$T_1$ and $T_2$ are optimal BSTs of $F$ and $H$,
respectively, then $(x ~ T_1 ~ T_2)$ is an optimal BST of $F
+_{x} H$. Moreover, the optimal BSTs of $F +_{x} H$ have
size equal to the sum of the size of the optimal search
trees of $F$ and $H$ plus one. Another simple consequence of
this theorem is that, if $x$ is a variable not in $F$, then
$x$ is an optimal backtracking branching variable of $F +_x
\bot = \{x\} \cup (\neg x \vee F)$.

We define the {\em product} of two sets of clauses as follows.

\begin{definition}

The product of two sets of clauses $F$ and $H$ is:
\eatpar

\[
F \cdot H = \{ \gamma \vee \delta ~|~
\gamma \in F \mbox{ and } \delta \in H \}
\]

\end{definition}

In the following, we will only consider the product of
formulae not sharing variables. In this case, $F \cdot H$ is
unsatisfiable if and only if both $F$ and $H$ are
unsatisfiable.

\begin{lemma}
\label{mult-struct}

If $F$ and $H$ do not share variables and are both
unsatisfiable, the tree obtained by replacing every empty
subtree of an optimal BST of $F$ with an optimal BST of $H$
is an optimal BST of $F \cdot H$.

\end{lemma}

\proof The claim is proved by induction on the total number
of variables of $F$ and $H$. The base case is when the total
number of variables of $F$ and $H$ is zero. Since both these
formulae are unsatisfiable, they are both only composed of
the empty clause. Their product is composed by the empty
clause only as well, and $()$ is its only BST.

Let us now assume that $F$ is built over $n$ variables while
$H$ is built over $m$ variables. We prove the claim
assuming that it holds for any pair of formulae whose total
number of variables is $n+m-1$.

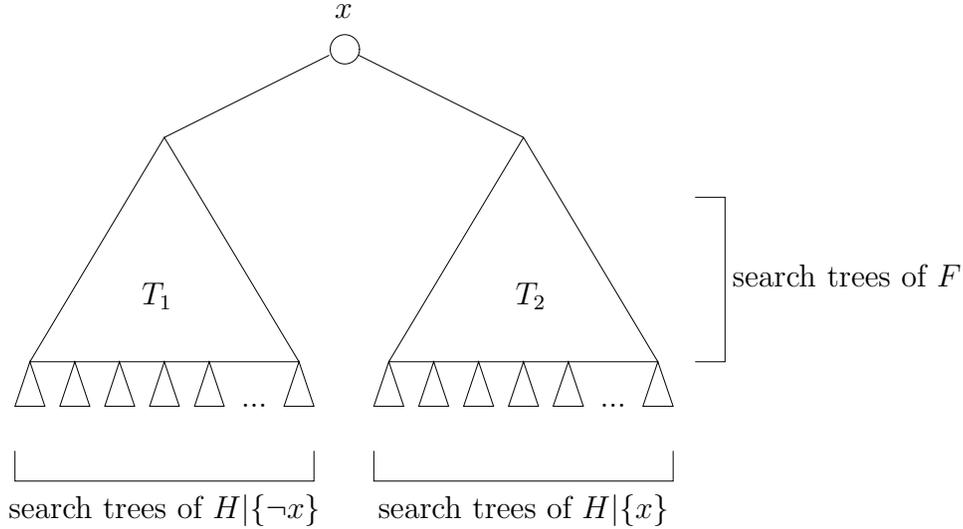
\begin{figure}[t]
\centering
\setlength{\unitlength}{0.00087489in}
\begingroup\makeatletter\ifx\SetFigFont\undefined%
\gdef\SetFigFont#1#2#3#4#5{%
  \reset@font\fontsize{#1}{#2pt}%
  \fontfamily{#3}\fontseries{#4}\fontshape{#5}%
  \selectfont}%
\fi\endgroup%
{\renewcommand{\dashlinestretch}{30}
\begin{picture}(5839,3265)(0,-10)
\put(1997,2928){\ellipse{180}{180}}
\path(912,2398)(102,1048)(1722,1048)(912,2398)
\path(3072,2398)(2262,1048)(3882,1048)(3072,2398)
\path(912,2398)(1902,2893)
\path(3072,2398)(2082,2893)
\path(912,1048)(822,778)(1002,778)(912,1048)
\path(642,1048)(552,778)(732,778)(642,1048)
\path(372,1048)(282,778)(462,778)(372,1048)
\path(1182,1048)(1092,778)(1272,778)(1182,1048)
\path(1722,1048)(1632,778)(1812,778)(1722,1048)
\path(102,1048)(12,778)(192,778)(102,1048)
\path(2262,1048)(2172,778)(2352,778)(2262,1048)
\path(2532,1048)(2442,778)(2622,778)(2532,1048)
\path(2802,1048)(2712,778)(2892,778)(2802,1048)
\path(3342,1048)(3252,778)(3432,778)(3342,1048)
\path(3882,1048)(3792,778)(3972,778)(3882,1048)
\path(3072,1048)(2982,778)(3162,778)(3072,1048)
\path(2172,508)(2172,328)(3972,328)(3972,508)
\path(12,508)(12,328)(1812,328)(1812,508)
\path(4107,2038)(4287,2038)(4287,1048)(4107,1048)
\put(867,1363){\makebox(0,0)[b]{{\SetFigFont{12}{14.4}{\familydefault}{\mddefault}{\updefault}$T_1$}}}
\put(3117,1363){\makebox(0,0)[b]{{\SetFigFont{12}{14.4}{\familydefault}{\mddefault}{\updefault}$T_2$}}}
\put(1992,3118){\makebox(0,0)[b]{{\SetFigFont{12}{14.4}{\familydefault}{\mddefault}{\updefault}$x$}}}
\put(3072,58){\makebox(0,0)[b]{{\SetFigFont{12}{14.4}{\familydefault}{\mddefault}{\updefault}search trees of $H|\{x\}$}}}
\put(912,58){\makebox(0,0)[b]{{\SetFigFont{12}{14.4}{\familydefault}{\mddefault}{\updefault}search trees of $H|\{ \neg x\}$}}}
\put(4332,1498){\makebox(0,0)[lb]{{\SetFigFont{12}{14.4}{\familydefault}{\mddefault}{\updefault}search trees of $F$}}}
\put(3612,778){\makebox(0,0)[b]{{\SetFigFont{12}{14.4}{\familydefault}{\mddefault}{\updefault}...}}}
\put(1452,778){\makebox(0,0)[b]{{\SetFigFont{12}{14.4}{\familydefault}{\mddefault}{\updefault}...}}}
\end{picture}
}
 %
%
\caption{Search tree of $F \cdot H$.}
\label{fig-mult}
\end{figure}

Let $(x~T_1'~T_2')$ be an optimal BST of $F \cdot H$. We
first consider the case in which $x$ is a variable of $F$
and then the case in which it is a variable of $H$. In both
cases, we show that this tree can be modified, without
changing its size, in such a way it satisfies the statement
of the theorem.

If $x$ is a variable of $F$, the two subtrees $T_1'$ and
$T_2'$ are optimal BSTs of $(F|\{\neg x\}) \cdot H$ and of
$(F|\{x\}) \cdot H$, respectively, because $x$ is not a
variable of $H$. As a result, they have the same size of any
other pair of optimal BSTs of these two formulae. In
particular, since these two formulae have $n+m-1$ variables,
they have two BSTs $T_1$ and $T_2$ that are as specified in
the statement of the theorem, \ie, $T_1$ is a tree of
$F|\{x\}$ where all empty subtrees are replaced with optimal
BSTs of $H$, and the same for $T_2$. The tree $(x~T_1~T_2)$
is therefore a tree that satisfies the condition in the
statement of the theorem. Note that, if the variables of $F$
and $H$ are not disjoint, what results by this construction
is an optimal BST of $F$ whose empty subtrees are replaced
by optimal BSTs of $H|\{\neg x\}$ and of $H|\{x\}$, instead
of optimal BSTs of $H$.

Let us now assume that $x$ is a variable of $H$. By
definition of BSTs, $T_1'$ and $T_2'$ are optimal BSTs of
$(F \cdot H)|\{\neg x\}$ and of $(F \cdot H)|\{x\}$,
respectively, which are the same as $F \cdot (H|\{\neg x\})$
and $F \cdot (H|\{x\})$, respectively. Since these two
formulae contain $n+m-1$ variables, they have two BSTs $T_1$
and $T_2$ that are as specified in the statement of the
theorem. Since $T_1$ and $T_2$ have the same size of $T_1'$
and $T_2'$, respectively, the tree $T=(x~T_1~T_2)$ is an
optimal BST of $F \cdot H$. This tree is as in
Figure~\ref{fig-mult}.

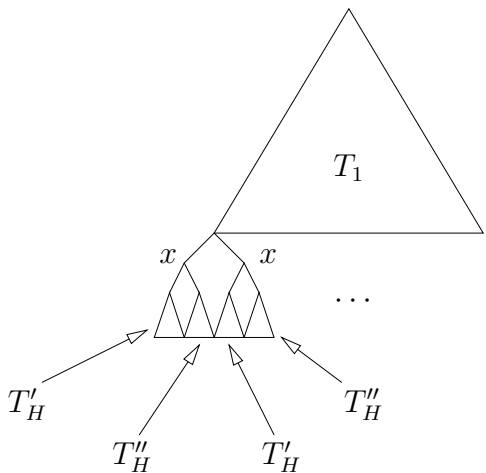
\begin{figure}[t]
\centering
\setlength{\unitlength}{0.00087489in}
\begingroup\makeatletter\ifx\SetFigFont\undefined%
\gdef\SetFigFont#1#2#3#4#5{%
  \reset@font\fontsize{#1}{#2pt}%
  \fontfamily{#3}\fontseries{#4}\fontshape{#5}%
  \selectfont}%
\fi\endgroup%
{\renewcommand{\dashlinestretch}{30}
\begin{picture}(3034,2875)(0,-10)
\path(2212,2848)(1402,1498)(3022,1498)(2212,2848)
\path(1402,1498)(1222,1318)
\path(1402,1498)(1582,1318)
\path(1222,1318)(1132,1138)
\path(1222,1318)(1312,1138)
\path(1312,1138)(1222,868)(1402,868)(1312,1138)
\path(1492,1138)(1402,868)(1582,868)(1492,1138)
\path(1672,1138)(1582,868)(1762,868)(1672,1138)
\path(1132,1138)(1042,868)(1222,868)(1132,1138)
\path(1582,1318)(1492,1138)
\path(1582,1318)(1672,1138)
\path(367,598)(997,913)
\whiten\path(903.085,832.502)(997.000,913.000)(876.252,886.167)(903.085,832.502)
\path(952,283)(1312,823)
\whiten\path(1270.397,706.513)(1312.000,823.000)(1220.474,739.795)(1270.397,706.513)
\path(1762,283)(1492,823)
\whiten\path(1572.498,729.085)(1492.000,823.000)(1518.833,702.252)(1572.498,729.085)
\path(2167,598)(1807,868)
\whiten\path(1921.000,820.000)(1807.000,868.000)(1885.000,772.000)(1921.000,820.000)
\put(2122,1048){\makebox(0,0)[lb]{{\SetFigFont{12}{14.4}{\familydefault}{\mddefault}{\updefault}$\cdots$}}}
\put(1177,1318){\makebox(0,0)[rb]{{\SetFigFont{12}{14.4}{\familydefault}{\mddefault}{\updefault}$x$}}}
\put(1672,1318){\makebox(0,0)[lb]{{\SetFigFont{12}{14.4}{\familydefault}{\mddefault}{\updefault}$x$}}}
\put(277,373){\makebox(0,0)[b]{{\SetFigFont{12}{14.4}{\familydefault}{\mddefault}{\updefault}$T_H'$}}}
\put(907,58){\makebox(0,0)[b]{{\SetFigFont{12}{14.4}{\familydefault}{\mddefault}{\updefault}$T_H''$}}}
\put(1807,58){\makebox(0,0)[b]{{\SetFigFont{12}{14.4}{\familydefault}{\mddefault}{\updefault}$T_H'$}}}
\put(2302,373){\makebox(0,0)[b]{{\SetFigFont{12}{14.4}{\familydefault}{\mddefault}{\updefault}$T_H''$}}}
\put(2212,1813){\makebox(0,0)[b]{{\SetFigFont{12}{14.4}{\familydefault}{\mddefault}{\updefault}$T_1$}}}
\end{picture}
}
 %
%
\caption{The rearrangement of the search tree of $F \cdot H$.}
\label{fig-mult-after}
\end{figure}

The optimal BSTs of $H|\{\neg x\}$ need not to be the same.
However, they have all the same size. As a result, they can
all be replaced by the same one $T_H'$. For the same reason,
the optimal BSTs of $H|\{x\}$ can all be replaced by the
same one $T_H''$. Since $x$ is not a variable of $F$, the
trees $T_1$ and $T_2$ are both search trees of $F$, and can
therefore be replaced by $T_1$. The resulting tree can be
rearranged by adding a number of copies of $x$ below $T_1$,
as shown in Figure~\ref{fig-mult-after}.

This tree is exactly as specified by the statement of the
theorem, and has been obtained from an optimal BST with
transformations that do not modify the size.~\qed

A consequence of this lemma is that the size of the
backtracking optimal search trees of $F \cdot H$ is equal to
the product of the size of the optimal search trees of $F$
and $H$, plus their sum.

The following corollary summarizes the results obtained so
far.

\begin{corollary}

There exists a constant $c$, where $0<c<1$, such that for
every positive integer $m$, there exists a set of clauses
$H_m$ such that\eatpar

\[
s(H_m) \in \{2^{cm},2^{cm}+1, \ldots, 2^m\}
\]

If $F$ and $H$ are two sets of clauses not sharing
variables, then:\eatpar

\begin{eqnarray*}
s(F \cup H) &=& \min(s(F), s(H))
\end{eqnarray*}

If $F$ and $H$ do not share variables and are both
unsatisfiable, then:\eatpar

\begin{eqnarray*}
s(F +_x H) &=& s(F) + s(H) +1 \\
s(F \cdot H) &=& s(F)s(H)+s(F)+s(H)
\end{eqnarray*}

\end{corollary}

 %

\section{DPLL and Backtracking}
\label{sec-dpll-back}

In this section, we first show that the results about how to
combine formulae allow improving the current results on the
complexity of choosing the branching literal in DPLL
(\np-hardness and \conp-hardness \cite{libe-00}.) We then
turn to the problem of search tree size.

\begin{theorem}
\label{branch-dlog}

The OBV problem for backtracking is \Dlog{2}-hard.

\end{theorem}

\proof The reduction is from the problem
\paritysat: given a sequence of formulae
$\{F_1, \ldots, F_r\}$, decide whether the first
unsatisfiable formula of the sequence is of odd index;
this problem is \Dlog{2}-hard \cite{wagn-90}.
We make the following simplifying assumptions, which
do not affect the complexity of this problem:

\begin{enumerate}
\itemsep=-\parsep

\item $r$ is even;

\item each formula is built over its own alphabet of $n$
variables;

\item both $F_{r-1}$ and $F_r$ are unsatisfiable.

\end{enumerate}

We translate the sequence $\{F_1,\ldots,F_r\}$ into the
set of clauses $F$ below.\eatpar

\begin{eqnarray*}
F &=& G \cup D \\
G &=& \bot +_x \\
&& (F_1 \cup ( H_m + H_m + \\
&& (F_3 \cup ( H_m + H_m + \\
&& \vdots \\
&& (F_{r-3} \cup ( H_m + H_m + \\
&& (F_{r-1}) \cdots ) \\
D &=& H_m + \\
&& (F_2 \cup (H_m + H_m + \\
&& (F_4 \cup (H_m + H_m + \\
&& \vdots \\
&& (F_{r-2} \cup (H_m + H_m + \\
&& (F_r) \cdots )
\end{eqnarray*}

In this definition, $m=2n/c$, where $c$ is the constant of
Lemma~\ref{exp}. We neglect the fact that $m$ should be a
square number, and assume that each $H_m$ is built over a
private set of variables. Formula $F$ can be built in time
polynomial in the size of the original instance of
\paritysat, as unions and sums only increase size of a
constant amount.

Since $F$ is the union of two sets of clauses $G$ and $D$
not sharing variables, its optimal BSTs are the minimal ones
among those of $G$ and those of $D$. By
Lemma~\ref{add-variable}, $x$ is an optimal branching
variable of $G$. It is therefore an optimal variable of $F$
if and only if $s(G) \leq s(D)$. What is left to prove is
that $s(G)$ is less than or equal to $s(D)$ if and only if
the first unsatisfiable formula of the sequence has odd
index.

Let $i$ be the index of the first unsatisfiable formula of
the subsequence of $\{F_1,\ldots,F_r\}$ composed only of the
formulae of odd index, and $j$ the same for the even
indexes. The values of $s(G)$ and $s(D)$ are:\eatpar

\begin{eqnarray*}
s(G) &=& 1 + \frac{i-1}2 (2s(H_m)+2) + s(F_i) \\
s(D) &=& s(H_m) + 1 + \frac{j-2}2(2s(H_m)+2) + s(F_j)
\end{eqnarray*}

These equations can be proved by observing that $D$ can be
rewritten as $H_m + (F_2 \cup (H_m + D'))$, where $D'$ is
the formula corresponding to the sequence
$\{F_3,\ldots,F_r\}$. A similar recursive definition can be
given for $G'$, where $G=\bot+_x G'$, \ie, $G'=F_1 \cup (H_m
+ H_m + G'')$ and $G''$ is the formula corresponding to
$\{F_3,\ldots,F_r\}$. The equations above can be verified
against the recursive definitions of $D$ and $G$.

Let us now assume that $i<j$. We have that
$i \leq j-1$, therefore:\eatpar

\begin{eqnarray*}
s(G)
&\leq&
1 + \frac{j-2}2 (2s(H_m)+2) + s(F_i)
\\
&=&
s(D) - s(F_j) + s(F_i) - s(H_m)
\\
&<&
s(D)
\end{eqnarray*}

The last step can be done because we have set $m$ in such
a way $s(H_m) > s(F_i)$ for any formula $s(F_i)$ of the
sequence. If $j<i$ we have $j \leq i-1$; therefore:\eatpar

\begin{eqnarray*}
s(D)
&\leq&
s(H_m) + 1 + \frac{i-3}2(2s(H_m)+2) + s(F_j)
\\
&=&
s(H_m) + 1 + \frac{i-1}2(2s(H_m)+2) - 2s(H_m) - 2 + s(F_j)
\\
&=&
s(G) - s(F_i) -s(H_m) - 2 + s(F_j)
\\
&<&
s(G)
\end{eqnarray*}

Since $x$ is optimal if and only if $s(G) \leq s(D)$, the
claim is proved.~\qed

By Lemma~\ref{dpll-back}, for any formula $F$ we can
determine (in polynomial time) a formula $G$ such that the
BSTs of $F$ corresponds to the DSTs of $G$. Replacing the
formula $F$ with the corresponding formula $G$ in the proof
above, we obtain a proof of \Dlog{2}-hardness of OBV for
DPLL.

\begin{corollary}

The problem OBV is \Dlog{2}-hard for DPLL.

\end{corollary}

 %

Let us now consider the OTS problem. This is the problem of
deciding whether a formula has a DPLL proof of size bounded
by a number $k$. This problem has been proved in \np\  by
Buss~\cite{buss-95} by showing a nondeterministic Turing machine
that works in pseudo-polynomial time. The problem is
therefore in \np\  only assuming that the size $k$ of the
required proof is expressed in unary notation. In fact, we
prove that the problem is harder, but we need formulae whose
optimal tree size is exponential. This is why the following
\conp-hardness result does not contrast with the proof of
membership to \np.

\begin{theorem}
\label{size-conp}

The problem OTS is \conp-hard for DPLL and backtracking.

\end{theorem}

\proof We prove that, given a formula $G$, its
unsatisfiability is equivalent to the existence of a BST, of
size bounded by $k$, for a formula $F$, where $F$ and $k$
can be computed from $G$ in polynomial time.

Namely, $F=G \cup H_m$, where $m=(n+1)/c$, and $k=2^n$,
where $n$ is the number of variables of $G$. If $G$ is
satisfiable, then the optimal BSTs of $F$ are those of $H_m$
and therefore $s(F)=s(H_m) \geq 2^{cm}=2^{n+1}$, which is
greater than $k$. If $G$ is unsatisfiable, it has BSTs of
size bounded by $2^n$. These trees are smaller than those of
$H_m$. As a result, the optimal trees of $F$ are the optimal
trees of $G$, whose size is bounded by $k$.

This result also holds for DPLL thanks to
Lemma~\ref{dpll-back}.~\qed

The problem being both \conp-hard and \np-hard
\cite{iwam-97} suggests it may be \Dp-hard. In this paper,
we show a proof of \Dp-hardness that however relies on the
existence of formulae whose optimal BST size is known
exactly and is an exponential in the size of the formula.

\begin{conjecture}[Exponential Exact Formulae]
\label{conjecture-exact}

There exists a polynomial-time algorithm that takes an
integer $m$ in unary notation and gives a formula $L_m$
whose optimal BSTs have size equal to $2^m$.

\end{conjecture}

The validity of this conjecture would allow building, in
polynomial time, a formula whose optimal BST size is $k$
even if $k$ is not a power of two. This formula $F$ can be
built by incrementally as follows:

\begin{enumerate}

\item start with $F=\bot$;

\item if $s(F)=k$, output $F$ and stop;

\item set $F$ to $F + L_m$, where $m$ is the maximal value
such that $s(F) + 2^m + 1 \leq k$; 

\item go to Point 2.

\end{enumerate}

In a logarithmic number of steps, we end up with a formula
whose optimal BSTs are of size $k$.

\begin{theorem}
 
If the Exponential Exact Formulae Conjecture is true,
then the problem OTS is \Dp-hard for DPLL and backtracking.

\end{theorem}

\proof This theorem is proved by combining the formulae used
in the proofs of \np-hardness and \conp-hardness in a single
one. Iwama~\cite{iwam-97} proved that the problem of checking
whether an unsatisfiable formula has a tree-like resolution
proof of bounded size is \np-hard. Since tree-like optimal
resolution proofs are also optimal backtracking proofs and
vice versa, this is also a proof of \np-hardness for
backtracking. A minor technical difference is that the size
of the proof is defined to be the total number of literals
in Iwama's proof; however, his result still holds if the
size of the proof is defined to be the number of nodes.

Since the problem is \np-hard, there exist two
polynomial-time functions $\alpha$ and $\beta$ such that a
formula $F$ is satisfiable if and only if the unsatisfiable
formula $\alpha(F)$ has search trees of size bounded by the
integer $\beta(F)$.

We use the problem sat/unsat: given two formulae, decide
whether the first is satisfiable but the second is not. Two
formulae $F$ and $E$ are in sat/unsat if and only if the
formula $D$ has search trees of size bounded by the number
$k$.\eatpar

\begin{eqnarray*}
D &=& ( (\alpha(F) \cdot L_r) + E ) \cup L_m \\
k &=& \beta(F) \cdot 2^r + \beta(F) + 2^r + 1 + 2^n 
\end{eqnarray*}

The numbers $r$ and $m$ are defined as $r = n+1$ and
$m=2\log k$, where $n$ is the number of variables of $E$.

Let us first assume that $E$ is satisfiable. Since
$(\alpha(F) \cdot L_r) + E$ is satisfiable in this
case, the optimal search trees of $D$ are exactly those of
$L_m$. Therefore, $s(D)=s(L_m)=2^m>k$.

Let us now assume that $E$ is unsatisfiable. Since $E$
contains $n$ variables, we have $s(E) \leq 2^n - 1 \leq k$.
If $F$ is satisfiable, we have $s(\alpha(F)) \leq \beta(F)$.
As a result,
$s((\alpha(F) \cdot L_r) + E ) \leq \beta(F) \cdot
2^r + \beta(F) + 2^r + 1 + 2^n - 1 < k$, which implies
$s(D) < k$.

If $E$ is unsatisfiable and $F$ is unsatisfiable as well, we
have $s(\alpha(F)) > \beta(F)$. As a result,
$s(\alpha(F)) \geq \beta(F)+1$, which implies
$s(\alpha(F) \cdot L_r) \geq
(\beta(F)+1) \cdot 2^r + (\beta(F)+1) + 2^r =
\beta(F) \cdot 2^r + 2^r + \beta(F)+1 +2^r > k$.

We have therefore proved that $E$ is unsatisfiable and $F$
is satisfiable if and only if $s(D) \leq k$. This proves
that the OTS problem is \Dp-hard. By Lemma~\ref{dpll-back},
the same complexity result holds for DPLL.~\qed

This hardness result can be used as an intermediate step for
the proof of a more precise complexity characterization of
the OTS problem.

\begin{theorem}

If the Exponential Exact Formulae Assumption is true,
the problem OTS is \Dlog{2}-hard for DPLL and backtracking.

\end{theorem}

\proof As a consequence of the last theorem, there exists a
pair of polynomial-time functions $\alpha$ and $\beta$ such
that $F$ is satisfiable and $G$ is unsatisfiable if and only
if $s(\alpha(F,G)) \leq \beta(F,G)$. We use these two
functions for showing that \paritysat\  can be reduced to
the problem of search tree size for backtracking.

Given a set of formulae $\{F_1, \ldots , F_r\}$, each built
over its private set of variables, the question of whether
the first unsatisfiable formula has odd index has positive
answer if either $F_1$ is unsatisfiable, or $F_1 \wedge F_2$
is satisfiable and $F_3$ is unsatisfiable, or $F_1 \wedge
\cdots \wedge F_4$ is satisfiable and $F_5$ is
unsatisfiable, etc.  This question can be expressed as an
OTS problem as follows.\eatpar

\begin{eqnarray*}
D &=& (\alpha(\true, F_1) + G_1) \cup
(\alpha(F_1 \wedge F_2, F_3 ) + G_3) \cup
(\alpha(F_1 \wedge \cdots \wedge F_4, F_5) +G_5) \cup
\cdots
\\
k &=& max(\{
\beta(\true, F_1),
\beta(F_1 \wedge F_2, F_3),
\beta(F_1 \wedge \cdots \wedge F_4, F_5),
\ldots
\}) + 1
\end{eqnarray*}

\noindent where $G_i$ is the formula obtained by adding a
number of formulae $L_m$ in such a way
$s(G_i) = k-1-\beta(F_1 \wedge \cdots \wedge F_{i-1}, F_i)$.

Let us first assume that the index $i$ of the first
unsatisfiable formula of the sequence $F_1,\ldots,F_r$ is
odd. We have:

\[
s(\alpha(F_1 \wedge \cdots \wedge F_{i-1}, F_i))
\leq
\beta(F_1 \wedge \cdots \wedge F_{i-1}, F_i)
\]

Since $s(G_i) = k-1-\beta(F_1 \wedge \cdots \wedge F_{i-1},
F_i)$, then $s(\alpha(F_1 \wedge \cdots \wedge F_{i-1},F_i)
+ G_i) = s(\alpha(F_1 \wedge \cdots \wedge F_{i-1},F_i)) +
s(G_i) +1 \leq k$. Since $D$ is a union that contains a term
whose proof size is less than or equal to $k$, the proof
size of $D$ (being the minimal among its terms) is less than
or equal to $k$.

Let us now instead assume that the first unsatisfiable
formula of the sequence is of even index. In this case, for
every odd index $i$, either $F_1 \wedge \cdots \wedge
F_{i-1}$ is unsatisfiable, or $F_i$ is satisfiable. As a
result, we have:

\[
s(\alpha(F_1 \wedge \cdots \wedge F_{i-1}, F_i))
>
\beta(F_1 \wedge \cdots \wedge F_{i-1},F_i)
\]

As a result, $s(\alpha(F_1 \wedge \cdots \wedge F_{i-1},
F_i) +G_i) > k$ for every odd index $i$. Since all parts of
$D$ have optimal search tree size greater than $k$, the
proofs of $D$ all have size greater than $k$. As an
immediate consequence of Lemma~\ref{dpll-back}, the same
complexity result holds for DPLL.~\qed

 %

 %

\section{Restricted-Branching DPLL}
\label{restricted}

Satisfiability provers are often used for solving real-world
problems that can be reduced to the problem of
satisfiability. Formulae produced this way often contain
variables whose value can be uniquely determined from the
values of the other ones. If branching is not allowed on
these variables, DPLL not only remains a complete
satisfiability algorithm, but is even made more efficient in
most cases
\cite{giun-etal-98,copt-etal-01,giun-seba-99,giun-etal-02,shtr-00}
(but not always \cite{craw-bake-94}.)

Backtracking is incomplete if we cannot branch over all
variables. However, the algorithm obtained by adding unit
propagation to backtracking (or, equivalently, deleting the
monotone literal rule from DPLL) is complete as DPLL is. We
call DPLL-Mono this algorithm. The search trees it generates
are called DPLL-Mono search trees, and abbreviated DMST. The
following theorem relates the search trees of DPLL and of
DPLL-Mono.

\begin{lemma}
\label{dpll-backup}

Let $F$ be a formula over variables $\{x_1,\ldots,x_n\}$,
and let $G$ be defined as follows:

\[
G = \{x_i \vee \neg y_i, \neg x_i \vee y_i\} \cup F
\]

\noindent Any DST of $G$ can be transformed into a DMST of
$F$ by replacing each $y_i$ with $x_i$.

\end{lemma}

\proof The monotone literal rule cannot be used on $G$
because all variables occur both positive and negative. We
have to prove that the same happens for any partial
assignment. Given an assignment, the value of $x_i$ can be
inferred by the monotone literal rule only if one clause
between $x_i \vee \neg y_i$ and $\neg x_i \vee y_i$ is
satisfied. This can only happen when either $x_i$ or $y_i$
are set to a value; if this is the case, unit propagation
assigns a value to the other one. As a result, the monotone
literal rule cannot be applied on $G$, making its DSTs
exactly the same as its DMSTs, which are in turn equivalent
to the DMSTs of $F$.~\qed

The next result we prove is that a formula can be modified
in such a way we can obtain an optimal search tree by
branching first on a subset of its variables of our choice.

\begin{definition}

Let $F=\{\gamma_1,\ldots,\gamma_m\}$ be a formula over a set
of variables $X \cup Y \cup Z$, such that the value of $Z$
can be obtained from any truth evaluation of $X \cup Y$ by
applying unit propagation in $F$. Let
$X=\{x_1,\ldots,x_n\}$. We define $c_X(F)$ as follows:
\eatpar

\begin{eqnarray*}
c_X(F) 
&=&
\{ \gamma_i \vee \neg a \vee \neg b ~|~ \gamma_i \in F \} \cup \\
&&
\{ \neg x_i \vee v_i ~|~ x_i \in X \} \cup
\{ x_i \vee v_i ~|~ x_i \in X \} \cup \\
&&
\{ \neg v_1 \vee \cdots \vee \neg v_n \vee a \} \cup
\{ \neg v_1 \vee \cdots \vee \neg v_n \vee b \}
\end{eqnarray*}

\noindent where $a$, $b$, and $\{v_1,\ldots,v_n\}$ are new
variables not appearing in $F$.

\end{definition}

Once the values of  $X \cup Y$ are determined, $v_1, \ldots,
v_n$ are set to \true\   by unit propagation because of $x_i
\vee v_i$ and $\neg x_i \vee v_i$; the variables of $a$ and
$b$ are set to \true\  by unit propagation because of $\neg
v_1 \vee \cdots \vee \neg v_n \vee a$ and $\neg v_1 \vee
\cdots \vee \neg v_n \vee b$. Simplifying $c_X(F)$ with
these values we obtain $F$. At this point, unit propagation
sets the values of $Z$ by assumption. We can therefore
conclude that $F$ is satisfiable if and only if $c_X(F)$ is.
Moreover, if $F$ is unsatisfiable, then restricting
branching on $X \cup Y$ still allows DPLL-Mono to prove that
$c_X(F)$ is unsatisfiable.

What is interesting about $c_X(F)$ is that some optimal
DMSTs of it are obtained by branching on the variables $X$
before those of $Y$.

\begin{theorem}
\label{circuit-two-sets}

Let $F$ be an unsatisfiable formula over variables $X \cup Y
\cup Z$, such that the value of $Z$ can be obtained from
that of $X \cup Y$ by unit propagation. Restricting
branching on the variables in $X \cup Y$, there exists an
optimal DMST of $c_X(F)$ made of a complete tree over $X$
in which trees over $Y$ replace the empty subtrees.

\end{theorem}

\proof By induction on the number of variables of $X \cup
Y$. If $F$ contains no variable, the empty tree is an
optimal DMST of it, and the empty tree satisfies the
condition of the theorem. If $F$ contains one variable,
either it is a variable of $X$ or it is a variable of $Y$.
The second case is easy to deal with, as $c_X(F)=\{y_1 \vee
\neg a \vee \neg b, \neg y_1 \vee \neg a \vee \neg b, a,
b\}$, and the empty tree is again an optimal DMST of this
formula. If the only variable $F$ contains is $x_1 \in X$,
then $c_X(F)=\{x_1 \vee \neg a \vee \neg b, \neg x_1 \vee
\neg a \vee \neg b, \neg x_1 \vee v_1, x_1 \vee v_1, \neg
v_1 \vee a, \neg v_1 \vee b\}$. This formula cannot be
proved unsatisfiable just by applying unit propagation.
Since branching is allowed only on $x_1$, the tree
$(x_1~()~())$ is the only DMST of it. This tree satisfies
the conditions of the theorem.

Let us now prove the induction case. If the root of an
optimal DMST of $c_X(F)$ is $x_i$, then its left and right
subtrees are DMSTs of $c_X(F)|\{\neg x_i\}$ and of
$c_X(F)|\{x_i\}$, which are the same formulae as
$c_X(F|\{\neg x_i\})$ and $c_X(F|\{x_i\})$, respectively. By
the induction hypotheses, these subtrees obey the statement
of the theorem, and the claim is proved.

Let us now consider the case in which the root of an optimal
DMST of $c_X(F)$ is a variable $y_i$. If $c_X(F)$ does not
contain any variable $x_i$, the statement of the theorem is
true, as $c_X(F)$ is equal to $F$ after the propagation of
$a$ and $b$. If there are some variables $x_i$, setting the
value of $y_i$ does not have any consequence on the other
variables. We can therefore use the induction hypothesis:
both subtrees satisfy the condition of the theorem, as
$c_X(F)|\{\neg y_i\} = c_X(F|\{\neg y_i\})$ and
$c_X(F)|\{y_i\} = c_X(F|\{y_i\})$. 

\begin{figure}[ht]
\centering
\setlength{\unitlength}{0.00087489in}
\begingroup\makeatletter\ifx\SetFigFont\undefined%
\gdef\SetFigFont#1#2#3#4#5{%
  \reset@font\fontsize{#1}{#2pt}%
  \fontfamily{#3}\fontseries{#4}\fontshape{#5}%
  \selectfont}%
\fi\endgroup%
{\renewcommand{\dashlinestretch}{30}
\begin{picture}(3246,2271)(0,-10)
\path(1158,513)(1068,333)
\path(1203,513)(1293,333)
\path(1068,333)(1293,333)
\path(393,513)(303,333)
\path(438,513)(528,333)
\path(303,333)(528,333)
\path(663,513)(573,333)
\path(708,513)(798,333)
\path(573,333)(798,333)
\path(2013,513)(1923,333)
\path(2058,513)(2148,333)
\path(1923,333)(2148,333)
\path(2283,513)(2193,333)
\path(2328,513)(2418,333)
\path(2193,333)(2418,333)
\path(2778,513)(2688,333)
\path(2823,513)(2913,333)
\path(2688,333)(2913,333)
\put(1598,1913){\ellipse{180}{180}}
\path(1518,1863)(798,1503)
\path(1698,1863)(2418,1503)
\path(798,1503)(303,513)(1293,513)(798,1503)
\path(2418,1503)(1923,513)(2913,513)(2418,1503)
\put(1608,2088){\makebox(0,0)[b]{{\SetFigFont{14}{16.8}{\familydefault}{\mddefault}{\updefault}$y_i$}}}
\put(798,693){\makebox(0,0)[b]{{\SetFigFont{14}{16.8}{\familydefault}{\mddefault}{\updefault}$T_1$}}}
\put(2418,693){\makebox(0,0)[b]{{\SetFigFont{14}{16.8}{\familydefault}{\mddefault}{\updefault}$T_2$}}}
\put(393,63){\makebox(0,0)[b]{{\SetFigFont{14}{16.8}{\familydefault}{\mddefault}{\updefault}$T_1^1$}}}
\put(663,63){\makebox(0,0)[b]{{\SetFigFont{14}{16.8}{\familydefault}{\mddefault}{\updefault}$T_1^2$}}}
\put(1968,63){\makebox(0,0)[b]{{\SetFigFont{14}{16.8}{\familydefault}{\mddefault}{\updefault}$T_2^1$}}}
\put(2328,63){\makebox(0,0)[b]{{\SetFigFont{14}{16.8}{\familydefault}{\mddefault}{\updefault}$T_2^2$}}}
\put(933,108){\makebox(0,0)[b]{{\SetFigFont{12}{14.4}{\familydefault}{\mddefault}{\updefault}...}}}
\put(2553,108){\makebox(0,0)[b]{{\SetFigFont{12}{14.4}{\familydefault}{\mddefault}{\updefault}...}}}
\put(1248,63){\makebox(0,0)[b]{{\SetFigFont{14}{16.8}{\familydefault}{\mddefault}{\updefault}$T_1^m$}}}
\put(2823,63){\makebox(0,0)[b]{{\SetFigFont{14}{16.8}{\familydefault}{\mddefault}{\updefault}$T_2^m$}}}
\end{picture}
}
 %
%
\caption{An optimal DMST of $F \cdot H$.}
\label{cs-duplicate}
\end{figure}
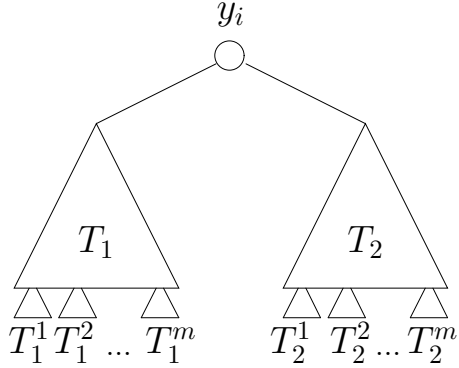

The tree $T$ is therefore as represented in
Figure~\ref{cs-duplicate}. This tree can be modified,
without changing neither its size nor the property of being
a search tree, as follows: replace $T_2$ with $T_1$. This is
possible because both trees are complete, so they have
exactly the same set of assignments at the leaves.
Therefore, by suitably changing the position of the trees on
$Y$ (\ie, $T_2^1$, $T_2^2$, $\ldots$, $T_2^m$) we obtain
another search tree, which has exactly the same size of the
original one.

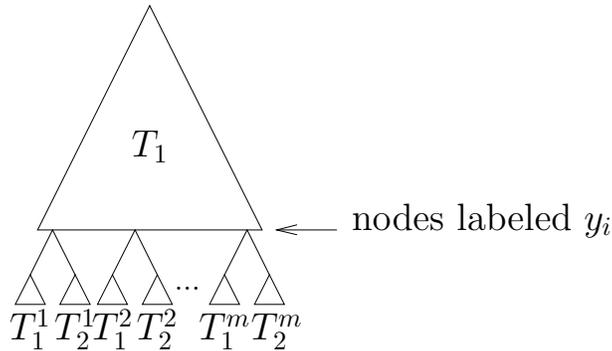
\begin{figure}[ht]
\centering
\setlength{\unitlength}{0.00087489in}
\begingroup\makeatletter\ifx\SetFigFont\undefined%
\gdef\SetFigFont#1#2#3#4#5{%
  \reset@font\fontsize{#1}{#2pt}%
  \fontfamily{#3}\fontseries{#4}\fontshape{#5}%
  \selectfont}%
\fi\endgroup%
{\renewcommand{\dashlinestretch}{30}
\begin{picture}(4051,2160)(0,-10)
\path(393,513)(303,333)(483,333)(393,513)
\path(663,513)(573,333)(753,333)(663,513)
\path(393,513)(528,783)(663,513)
\path(888,513)(798,333)(978,333)(888,513)
\path(1158,513)(1068,333)(1248,333)(1158,513)
\path(888,513)(1023,783)(1158,513)
\path(1563,513)(1473,333)(1653,333)(1563,513)
\path(1833,513)(1743,333)(1923,333)(1833,513)
\path(1563,513)(1698,783)(1833,513)
\path(438,783)(1788,783)(1113,2133)(438,783)
\path(2238,783)(1878,783)
\path(1998.000,813.000)(1878.000,783.000)(1998.000,753.000)
\put(2328,738){\makebox(0,0)[lb]{{\SetFigFont{14}{16.8}{\familydefault}{\mddefault}{\updefault}nodes labeled $y_i$}}}
\put(1338,423){\makebox(0,0)[b]{{\SetFigFont{12}{14.4}{\familydefault}{\mddefault}{\updefault}...}}}
\put(393,63){\makebox(0,0)[b]{{\SetFigFont{14}{16.8}{\familydefault}{\mddefault}{\updefault}$T_1^1$}}}
\put(1113,1188){\makebox(0,0)[b]{{\SetFigFont{14}{16.8}{\familydefault}{\mddefault}{\updefault}$T_1$}}}
\put(663,63){\makebox(0,0)[b]{{\SetFigFont{14}{16.8}{\familydefault}{\mddefault}{\updefault}$T_2^1$}}}
\put(888,63){\makebox(0,0)[b]{{\SetFigFont{14}{16.8}{\familydefault}{\mddefault}{\updefault}$T_1^2$}}}
\put(1158,63){\makebox(0,0)[b]{{\SetFigFont{14}{16.8}{\familydefault}{\mddefault}{\updefault}$T_2^2$}}}
\put(1563,63){\makebox(0,0)[b]{{\SetFigFont{14}{16.8}{\familydefault}{\mddefault}{\updefault}$T_1^m$}}}
\put(1878,63){\makebox(0,0)[b]{{\SetFigFont{14}{16.8}{\familydefault}{\mddefault}{\updefault}$T_2^m$}}}
\end{picture}
}
 %
%
\caption{The result of the transformation.}
\label{cx-final}
\end{figure}

Another step of the transformation is to replace
$(y_i~T_1~T_1)$ with the tree obtained by replacing each
empty subtree with $(y_i~()~())$ in $T_1$. This tree has
exactly the same size of the original one, and the same
assignments in the leaves. As a result, by adding the
subtrees $T_1^i$ and $T_2^i$ we still obtain a search tree,
which is shown in Figure~\ref{cx-final}.

This tree satisfies the condition of the theorem.~\qed

From the shape of the optimal DMST of $c_X(F)$, we can
infer their size.

\begin{corollary}
\label{circuit-two-sets-size}

Let $F$ be a formula over $X \cup Y \cup Z$, where $|X|=n$.
Assuming branching is allowed only on $X \cup Y$, we have:
\eatpar

\[
s(c_X(F)) = 2^{n} - 1 +
\sum_{
X' {\rm ~is~a~model~over~} X
} s(F|{X'})
\]

\end{corollary}

The previous theorem also shows that the sum of formulae can
be defined as $F +_x G = c_{\{x\}} ( (F \vee x) \cup (G \vee
\neg x) )$ for DPLL-Mono (this is useful, as it is not clear
whether Lemma~\ref{add-variable} holds for DPLL-Mono.) By
Theorem~\ref{circuit-two-sets}, indeed, there exists an
optimal search tree of $F +_x G$ containing $x$ in the root,
and the two subtrees are optimal DMST of $F$ and $G$,
respectively. As a result, the size of the optimal DPLL-Mono
search trees of $F +_x G$ is $s(F)+s(G)+1$.

The property about the union of two formulae $F \cup G$
still hold for backtracking with unit propagation (the
proof is like the one for backtracking). Formulae $H_m$
can be replaced with formulae whose optimal search trees
have an exact exponential value.

\begin{corollary}

The optimal DMST of $V_n=c_X(X \cup \{y,\neg y\})$, where
$X=\{x_1,\ldots,x_n\}$, have size $2^{n}-1$.

\end{corollary}

Formulae that have exact size can be built easily even if
the size is not equal to $2^n-1$ for some $n$: using the
same construction reported after
Conjecture~\ref{conjecture-exact}, we can build a formula
$I_m$ that has optimal DMST size equal to $m$ in polynomial
time, for every $m>0$. These formulae allow for reducing the
OTS problem to the OBV problem.

\begin{theorem}
\label{size-branch}

For restricted-branching DPLL-Mono, the OTS problem can be
polynomially reduced to the OBV problem.

\end{theorem}

\proof Given a formula $G$, we know that $s(G) \leq k$
if and only if $a$ is the optimal branching variable
of $(\bot +_a G) \cup I_{k+1}$.~\qed

Another consequence of Theorem~\ref{circuit-two-sets}
is the possibility of relating the search tree size of a
formula with the number of models of another one.

\begin{corollary}
\label{circuit-linear}

Let $G$ be a formula over $X$. Restricting branching over
the variables in $X \cup \{y\}$, where $y$ is a new variable
not in $X$, the size of the optimal DMSTs of $e_X(G)$ is
$2^{n+1} -1  +2 |\mod(G)|$, where $e_X(G)$ is defined as follows:

\[
e_X(G)=
c_X(G \cup
\{y,\neg y\})
\]

\end{corollary}

\def\minorsat{{\profont minsat}}
\def\majsat{{\profont majsat}}

\def\eminorsat{{\profont e-minsat}}
\def\emajsat{{\profont e-majsat}}

We prove that the problem of search tree size is hard for
the class \nppp. First of all, we need a complete problem
for this class. We use \eminorsat: given a formula $F$ over
variables $X \cup Y$, decide whether there exists a truth
assignment over $X$ such that at most half of the models
extending it satisfy $F$. The similar problem where ``at
most'' is replaced by ``at least'' is called \emajsat, and
is \nppp-complete \cite{litt-etal-98}. Proving that
\eminorsat\ is complete for the same class is an easy
exercise.

\begin{theorem}

Checking whether the size of the optimal DMST of a formula
is bounded by a number in binary notation is \nppp-hard for
restricted-branching DPLL-Mono.

\end{theorem}

\proof We reduce \eminorsat\  to the problem of search tree
size. Let $F$ be a formula over $X \cup Y$, where
$|X|=|Y|=n$. Given a truth evaluation $X'$ over $X$, the
number of models of $F|{X'}$ are related to the formula
$c_Y((F|{X'}) \cup I_1)$ by
Corollary~\ref{circuit-two-sets-size}:\eatpar

\begin{eqnarray*}
s(c_Y((F|{X'}) \cup I_1))
&=&
2^{n} - 1 + \sum_{Y' {\rm ~is~a~model~of~} Y} s((F|{X' \cup Y'}) \cup I_1)
\\
&=&
2^{n} - 1  + | \mod(F|{X'}) |
\end{eqnarray*}

The optimal tree size of $c_Y(F|{X'} \cup I_1)$ linearly
depends on the number of models of $F|{X'}$. The reduction
is completed by the addition of a formula $I_{k}$, where
$k=2^{n+1}+2^{n-1}$. Indeed, the formula $c_Y((F|{X'})
\cup I_1) \cup I_{k}$ has the following property: \eatpar

\begin{eqnarray*}
s(c_Y((F|{X'}) \cup I_1) \cup I_{k})
&<& k \mbox{ if $F|{X'}$ has at most half of the models}
\\
s(c_Y((F|{X'}) \cup I_1) \cup I_{k})
&=&
k \mbox{ otherwise}
\end{eqnarray*}

By combining Corollary~\ref{circuit-two-sets-size} with the
above inequalities, we obtain a way for summing up the size
of the search trees of all formulae $F|{X'}$. The optimal
search trees of $c_X(c_Y(F \cup I_1) \cup I_{k}))$ have
indeed the following size: \eatpar

\begin{eqnarray*}
s(c_X(c_Y(F \cup I_1) \cup I_{k})))
&=&
2^{n} - 1 + 2^n k
\mbox{ if all $F|{X'}$ have more than half models}
\\
&<&
2^{n} - 1  + 2^n k
\mbox{ otherwise}
\end{eqnarray*}

This proves that $c_X(c_Y(F \cup I_1) \cup I_{k}))$
has search trees of size bounded by $2^{n} + 2^n k -2$ if and
only if there exists $X'$ such that $F|{X'}$ has at most
half of the models.~\qed

Theorem~\ref{size-branch} shows that the OTS problem can be
polynomially reduced to the OBV problem. Moreover,
Lemma~\ref{dpll-backup} shows that any formula can be
translated into another one whose DST are the DMST of the
original one. This is therefore a reduction from the OTS
problem for DPLL-Mono to the OTS problem for DPLL.

\begin{corollary}

The problems OTS and OBV for restricted-branching DPLL are
\nppp-hard.

\end{corollary}

\nppp\  contains $\p^\pp$~\cite{toda-91}, which in turn
contains the whole polynomial hierarchy. As a result, the
above theorem shows that the problem of search tree size is
hard for any class of the polynomial hierarchy.

 %

\section{Regular Resolution}
\label{sec-resolution}

In this section, we consider the problems of the proof size
and of the optimal choice for regular resolution. We proceed
by first checking which results for backtracking and DPLL
continue to hold for regular resolution, and then proving
hardness results from them.
\nocite{bone-etal-00-b,bens-impa-widg-00} Formulae having
exponential optimal resolution proofs exist both for regular
and general resolution~\cite{tsei-70,hake-85,urqu-87}. The
result $s(F \cup H) = \min( s(F), s(H))$ holds for
resolution: since the clauses of $F$ and $H$ do not share
variables, resolution can only be applied between two
clauses of $F$ or between two clauses of $H$. This implies
that any optimal resolution proof either contains clauses of
$F$ only or of $H$ only. It is not clear whether the
properties of multiplication and sum hold for resolution.

\comment
Theorem 3(add-variable) may hold
\endcomment

The problem of proof size is \np-hard because of a result by
Iwama~\cite{iwam-97} (this result also holds if the size of a
resolution proof is defined to be the number of generated
clauses instead of the total number of literals.) Using the
union of formulae, we can prove that the problem is
\conp-hard as well: if $H_m$ is a formula whose optimal
proof size is greater than $2^n$, then the formula $G \cup
H_m$ have proof size less than or equal to $2^n$ if and only
if $G$ is unsatisfiable, where $G$ is a formula over $n$
variables.

\begin{theorem}

Deciding whether there exists a regular resolution proof of
a formula, of size bounded by a number, is \conp-hard.

\end{theorem}

Let us now consider the problem of the optimal choice, \ie,
whether two clauses are brother leaves of an optimal proof.
In order to prove a hardness result, we need a way for
building formulae for which an optimal choice is known.

\begin{lemma}

Let $F$ be an unsatisfiable formula such that $F \backslash
\{\gamma\}$ is satisfiable, $x$ a new variable not in
$F$, and $g_x(F)$ the following formula:

\[
g_x(F) = \{ x, \neg x \vee \gamma \} \cup
F \backslash \{\gamma\}
\]

\noindent All optimal regular resolution proofs of $g_x(F)$
contain exactly one resolution step involving $x$. Such a
step can be pushed to the leaves of the proof. 

\end{lemma}

\proof Since $\gamma$ is needed to make $F$ unsatisfiable,
the clauses $x$ and $\neg x \vee \gamma$ are both needed to
make $g_x(F)$ unsatisfiable. Therefore, they are both leaves
of any resolution proof of $g_x(F)$. We can also show that
some optimal proofs of $g_x(F)$ actually contain the
resolution of these two clauses.

\begin{figure}[ht]
\centering
\setlength{\unitlength}{0.00087489in}
\begingroup\makeatletter\ifx\SetFigFont\undefined%
\gdef\SetFigFont#1#2#3#4#5{%
  \reset@font\fontsize{#1}{#2pt}%
  \fontfamily{#3}\fontseries{#4}\fontshape{#5}%
  \selectfont}%
\fi\endgroup%
{\renewcommand{\dashlinestretch}{30}
\begin{picture}(3489,2695)(0,-10)
\put(2567,1143){\ellipse{180}{180}}
\put(2037,1138){\ellipse{180}{180}}
\put(1722,328){\ellipse{180}{180}}
\put(2303,1502){\ellipse{180}{180}}
\put(677,1038){\ellipse{180}{180}}
\put(899,1412){\ellipse{180}{180}}
\put(1132,1048){\ellipse{180}{180}}
\path(2085,1224)(2239,1430)
\path(1762,414)(2017,699)(1819,898)(1969,1070)
\path(2512,1209)(2359,1430)
\path(839,1337)(704,1112)
\path(1081,1124)(950,1330)
\path(946,1499)(1092,1720)
\path(597,328)(2937,328)(3477,1453)
	(1767,2668)(12,1318)(597,328)
\path(1902,553)(1227,1003)
\path(2352,1588)(2532,1858)
\put(1542,58){\makebox(0,0)[lb]{{\SetFigFont{12}{14.4}{\familydefault}{\mddefault}{\updefault}$\neg x \vee \gamma$}}}
\put(2712,1138){\makebox(0,0)[lb]{{\SetFigFont{12}{14.4}{\familydefault}{\mddefault}{\updefault}$x$}}}
\put(507,823){\makebox(0,0)[lb]{{\SetFigFont{12}{14.4}{\familydefault}{\mddefault}{\updefault}$x$}}}
\put(732,1543){\makebox(0,0)[lb]{{\SetFigFont{12}{14.4}{\familydefault}{\mddefault}{\updefault}$\delta_2$}}}
\put(2442,1453){\makebox(0,0)[lb]{{\SetFigFont{12}{14.4}{\familydefault}{\mddefault}{\updefault}$\delta_1$}}}
\put(1047,1183){\makebox(0,0)[lb]{{\SetFigFont{12}{14.4}{\familydefault}{\mddefault}{\updefault}$\neg x \vee \delta_2$}}}
\put(1902,868){\makebox(0,0)[lb]{{\SetFigFont{12}{14.4}{\familydefault}{\mddefault}{\updefault}$\neg x \vee \delta_1$}}}
\end{picture}
}
 %
%
\caption{A resolution proof of $g_x(F)$.}
\label{resol1}
\end{figure}
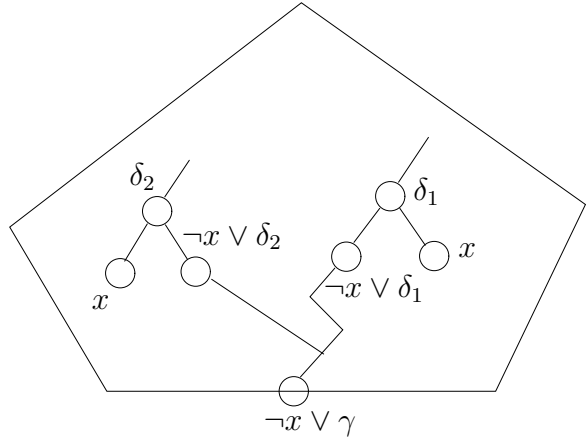

Since $\neg x \vee \gamma$ is a leaf of all regular
resolution proofs of $g_x(F)$ but the root of the proof does
not contain literals then, in any path from $\neg x \vee
\gamma$ to the root, there is a resolution step that
eliminates $\neg x$. The only clause that can eliminate
$\neg x$ is $x$. As a result, every path from $\neg x \vee
\gamma$ to the root contains the resolution of a clause
$\neg x \vee \delta$ with the clause $x$. Since the proof is
a DAG, there may be more than one such path. However, since
we assume we are using regular resolution, no path contains
more than one resolution with $x$. Figure~\ref{resol1} shows
an example of such a proof.

We show that, if $\neg x \vee \delta_1,\ldots, \neg x\vee
\delta_m$ are the clauses that are resolved with $x$, then
all these resolution steps can be replaced by the single
resolution of $x$ with $\neg x \vee \gamma$. This is
possible because, in the path from $\neg x \vee \gamma$ to
$\neg x \vee \delta_i$, the variable $\neg x$ is present in
all clauses (because this is a regular resolution proof).

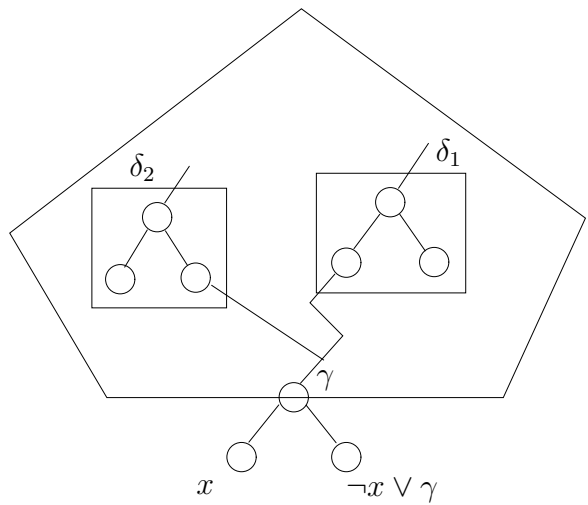
\begin{figure}[ht]
\centering
\setlength{\unitlength}{0.00087489in}
\begingroup\makeatletter\ifx\SetFigFont\undefined%
\gdef\SetFigFont#1#2#3#4#5{%
  \reset@font\fontsize{#1}{#2pt}%
  \fontfamily{#3}\fontseries{#4}\fontshape{#5}%
  \selectfont}%
\fi\endgroup%
{\renewcommand{\dashlinestretch}{30}
\begin{picture}(3489,3055)(0,-10)
\put(2567,1503){\ellipse{180}{180}}
\put(2037,1498){\ellipse{180}{180}}
\put(1722,688){\ellipse{180}{180}}
\put(2303,1862){\ellipse{180}{180}}
\put(677,1398){\ellipse{180}{180}}
\put(899,1772){\ellipse{180}{180}}
\put(1132,1408){\ellipse{180}{180}}
\put(1407,328){\ellipse{180}{180}}
\put(2037,328){\ellipse{180}{180}}
\path(2085,1584)(2239,1790)
\path(1762,774)(2017,1059)(1819,1258)(1969,1430)
\path(2512,1569)(2359,1790)
\path(839,1697)(704,1472)
\path(1081,1484)(950,1690)
\path(946,1859)(1092,2080)
\path(597,688)(2982,688)(3477,1768)
	(1767,3028)(12,1678)(597,688)
\path(1902,913)(1227,1363)
\path(2352,1948)(2532,2218)
\path(507,1948)(1317,1948)(1317,1228)
	(507,1228)(507,1948)
\path(1857,2038)(2757,2038)(2757,1318)
	(1857,1318)(1857,2038)
\path(1452,418)(1632,643)
\path(1976,415)(1796,640)
\put(2577,2083){\makebox(0,0)[lb]{{\SetFigFont{12}{14.4}{\familydefault}{\mddefault}{\updefault}$\delta_1$}}}
\put(2037,58){\makebox(0,0)[lb]{{\SetFigFont{12}{14.4}{\familydefault}{\mddefault}{\updefault}$\neg x \vee \gamma$}}}
\put(1137,103){\makebox(0,0)[lb]{{\SetFigFont{12}{14.4}{\familydefault}{\mddefault}{\updefault}$x$}}}
\put(1857,733){\makebox(0,0)[lb]{{\SetFigFont{12}{14.4}{\familydefault}{\mddefault}{\updefault}$\gamma$}}}
\put(732,1993){\makebox(0,0)[lb]{{\SetFigFont{12}{14.4}{\familydefault}{\mddefault}{\updefault}$\delta_2$}}}
\end{picture}
}
 %
%
\caption{Pushing down the resolution of $x$ and $\neg x \vee \gamma$.}
\label{resol2}
\end{figure}

The transformation is as follows: we first identify the
three nodes $\neg x \vee \delta_i$, $x$, and $\delta_i$ for
each $i$; we then remove all literals $\neg x$ from internal
nodes of the DAG; we then replace the leaf $\neg x \vee
\gamma$ with the resolution of $\neg x \vee \gamma$ and $x$.
This leads to a new regular resolution proof, made like the
one in Figure~\ref{resol2}.

If the number of clauses $\delta_i$ is greater than one, the
proof is made smaller, thus contradicting the assumption of
optimality. This proves that the optimal proofs only contain
one resolution step involving $x$. In this case, the size of
the proof is left unchanged by the transformation that
pushes the resolution of $x$ to the leaves of the DAG.~\qed

This lemma tells how to modify a formula in such a way an
initial resolution step is known, but it only holds when a
clause $\gamma$ is known to be necessary to make the formula
unsatisfiable. We now remove this assumption.

\begin{lemma}

If $F$ is an unsatisfiable formula not containing the
variables $x$ and $y$, all optimal regular resolution proof
of $f_y^x(F)=\{x, \neg x \vee y\} \cup \{\neg y \vee \delta
~|~ \delta \in F\}$ contain exactly one resolution of $x$,
which can be pushed to the leaf level.

\end{lemma}

\proof The unit clause $y$ is needed to make $\{\neg y \vee
\delta ~|~ \delta \in F\}$ unsatisfiable, if $F$ is
unsatisfiable. The previous lemma therefore applies.~\qed

The complexity of the optimal regular resolution pair is
characterized as follows.

\begin{theorem}

The ORP problem is both \np-hard and \conp-hard.

\end{theorem}

\proof Let $F$ be a formula on $n$ variables, and let $H_m$
be a formula whose optimal regular resolution proofs are
larger than $2^{n}$. We show that both the satisfiability
and the unsatisfiability of $F$ can be reduced to the ORP
problem. By the previous lemma, if $F$ is unsatisfiable then
the resolution of $x$ with $\neg x \vee y$ is an optimal
choice for $f^x_y(F)$. For the same reason, the resolution
of $w$ with $\neg w \vee z$ is optimal for $f^w_z(H_m)$.

Consider the formula $f^x_y(F) \cup f^w_z(H_m)$. Since
$f^x_y(F)$ and $f^w_z(H_m)$ do not share variables, every
resolution proof of it either contains only clauses of
$f^x_y(F)$ or only clauses of $f^w_z(H_m)$: otherwise, the
proof would not form a connected DAG. Since $H_m$ is
unsatisfiable, $f^w_z(H_m)$ is unsatisfiable as well.

On the other hand, the satisfiability of $f^x_y(F)$ depends
on that of $F$. If $F$ is satisfiable, then $f^x_y(F)$ is
satisfiable as well. As a result, the proofs of $f^x_y(F)
\cup f^w_z(H_m)$ are exactly the proofs of the only
unsatisfiable formula of the union, that is, $f^w_z(H_m)$.
Resolving $w$ and $\neg w \vee z$ is therefore an optimal
choice, while resolving $x$ and $\neg x \vee y$ is not.

If $F$ is unsatisfiable, so is $f^x_y(F)$. The optimal
proofs of $F$ are at most $2^n$ large. A proof for
$f^x_y(F)$ can be obtained by adding $y$ to all nodes of a
proof of $F$, and then resolving its root with the result of
the resolution of $x$ with $\neg x \vee y$. As a result,
$f^x_y(F)$ has a regular resolution tree of size $2^n+2$.

Since $H_m$ is unsatisfiable, so is $f^w_z(H_m)$. Any
regular resolution proof of this formula can be modified in
such a way the resolution of $w$ and $\neg w \vee z$ is at
the leaf level. The rest of the proof is a proof of $H_m$
with the addition of $z$ to all clauses (otherwise, $z$
would be resolved more than once, leading to a larger
proof.) As a result, any proof of $f^w_z(H_m)$ has size
greater than or equal than that of $H_m$ plus two.

The smaller between the proofs of $f^x_y(F)$ and of
$f^w_z(H_m)$ are the former ones. We can therefore conclude
that, if $F$ is unsatisfiable, then $x$ and $\neg x \vee y$
is an optimal choice for $f^x_y(F) \cup f^w_z(H_m)$, while
$w$ and $\neg w \vee z$ is not. Since we have already proved
the converse if $F$ is satisfiable, the claim is
proved.~\qed

 %

\section{Conclusions}
\label{sec-conclusions}

In this paper, we have enhanced two complexity results about
the complexity of DPLL and resolution: namely, the
complexity of choosing the best branching variables is not
only \np-hard and \conp-hard \cite{libe-00}, but is also
\Dlog{2}-hard; the problem of proof size is not only
\np-hard \cite{iwam-miya-95,iwam-97,alek-etal-01}, but also
\conp-hard, if the size bound is in binary notation.

The problem of the search tree size can be also proved to be
\Dlog{2}-hard by assuming the possibility of building, in
polynomial time, a formula whose optimal search tree size is
exactly known and exponential. While this seems likely, no
formal proof of it is in the literature. Namely, we known
how to build, in polynomial time, formulae of exponential
optimal search tree size, but only a lower bound of this
optimal size is known, not the exact value. The possibility
of building these formulae is also related to the similarity
of the problems of search tree size and that of optimal
choice: if this is the case, indeed, the problems of optimal
choice and optimal search tree size can be easily reduced to
each other.

Let us now compare with other work in the literature. The
problem of search tree size has been already analyzed for
various proof systems. For backtracking, this problem has
been shown \np-complete
\cite{iwam-miya-95,iwam-97,alek-etal-01}. The membership
into \np, however, only holds if the number $k$ of the
question ``is there any proof of size bounded by $k$?'' is
in unary notation. The intuitive meaning of using the unary
notation is that the proof to search for should be small
enough to be stored. The binary representation makes sense
either when the proof is represented in some succinct form,
or when we only want to evaluate the proof size (without
finding it). Most SAT checkers developed in AI, for example,
are not aimed at producing a proof of unsatisfiability, but
only at producing a correct answer.

A problem that is related to the complexity of choice and of
tree size is that of automatizability of proof systems. A
proof system is called automatizable if a proof can be
produced in time that is polynomial in that of the optimal
proofs (the generated proof can therefore only polynomially
larger than the optimal ones.) The problems of optimal
choice and automatizability, while somehow close to each
other, are however different. Automatizability is about the
time needed to generate the whole proof; the optimal choice
problem is that of making, at each step, the optimal choice.
The first question is ``global'', as it involves the whole
proof; the second one is ``local'', as it is about a single
step of the proof. Doing a single step may be hard, while
the other steps of the proof are easy: if this is the case,
automatizability may be feasible while the problem of the
optimal choice remains hard.

The relative importance of automatizability and optimal
choice complexity depends on the expected application of the
satisfiability algorithms. If a complete proof is required,
the running time of a satisfiability checker has to be
measured w.r.t.\ this output. Therefore, automatizability is
important. On the other hand, many applications
\cite{erns-mill-weld-97,mass-marr-00-JAR} require a proof
only if the formula is satisfiable (\ie, they require a
model if it exists.) Most algorithms used in this case are
oblivious: each choice is made neglecting the previous ones.
These algorithms face the choice of the branching variable.

Finally, let us discuss the questions left open. The
problems about DPLL are only known to be hard for classes at
the second level of the polynomial hierarchy, while the only
class they are known to belong to is \pspace. The results on
restricted branching are more precise, as the problem are
hard for all classes of the polynomial hierarchy.
Unrestricted branching may be as hard as restricted
branching, but no proof of this claim has been found.

A large gap between the hardness and the membership results
is present in our results for regular resolution. The method
used for proving that problems about DPLL are in \pspace\
does not work for resolution. The problem is that resolution
proofs are DAGs, not trees. Therefore, iteratively guessing
a choice and checking the total size does not work, as two
nodes of the DAG may have the same child. This argument is a
hint that neither the problem of the optimal choice nor that
of the size are in \pspace\  for resolution.

 %

\bibliographystyle{plain}

\end{document}